\documentclass[prl,twocolumn,showpacs,preprintnumbers]{revtex4}
\usepackage{graphicx}
\begin{document}

\title{Structure of end states for a Haldane Spin Chain}

\author{M. Kenzelmann$^{1,2}$, G. Xu,$^{3}$ I.~A. Zaliznyak,$^{3}$
C. Broholm$^{1,2}$, J.~F. DiTusa,$^{4}$ G. Aeppli,$^{5,6}$ T.
Ito,$^{7}$ K. Oka$^{7}$ and H. Takagi$^{8}$}

\affiliation{(1) Department of Physics and Astronomy, Johns
Hopkins University, Baltimore, Maryland 21218\\(2) NIST Center for
Neutron Research, National Institute of Standards and Technology,
Gaithersburg, Maryland 20899\\(3) Department of Physics,
Brookhaven National Laboratory, Upton, New York 11973-5000\\(4)
Department of Physics and Astronomy, Louisiana State University,
Baton Rouge, Louisiana 70803\\(5) London Centre for
Nanotechnology, Gower Street, London WC1E 6BT, UK\\ (6) NEC
Research Institute, 4 Independence Way, Princeton, New Jersey
08540\\(7) National Institute of Advanced Industrial Science and
Technology, Tsukuba, Ibaraki 305-8562, Japan\\(8) Graduate School
of Frontier Sciences, University of Tokyo, Hongo, Bukyo-ku, Tokyo
113-8656, Japan}
\date{\today}

\begin{abstract}
Inelastic neutron scattering was used to probe edge states in a
quantum spin liquid. The experiment was performed on finite length
antiferromagnetic  spin-1 chains in ${\rm
Y_2BaNi_{1-x}Mg_{x}O_5}$. At finite fields, there is a Zeeman
resonance below the Haldane gap. The wave vector dependence of its
intensity provides direct evidence for staggered magnetization at
chain ends, which decays exponentially towards the bulk ($\xi$ =
$8(1)$ at $T=0.1\;\mathrm{K}$). Continuum contributions to the
chain end spectrum indicate inter-chain-segment interactions. We
also observe a finite size blue shift of the Haldane gap.
\end{abstract}
\pacs{75.25.+z, 75.10.Jm, 75.40.Gb} \maketitle

Highly-correlated electron systems can have quantum ground states
that are coherent over macroscopic length scales. Among the best
known examples is the two-dimensional electron gas in a magnetic
field where Coulomb interactions lead to cooperative excitations
with fractional charge as evidenced in the quantum Hall effect.
Spin systems with strong quantum fluctuations feature similar
cooperative phenomena with coherent spin excitations that differ
fundamentally from classical spin waves.\par

Impurities and boundaries can liberate novel composite degrees of
freedom and dramatically change the physical properties of
macroscopic quantum systems. Spin correlations in the vicinity of
impurities are of particular interest because they reflect the
nature of the underlying bulk quantum state, and because they may
be important for understanding unconventional bulk properties.
Many recent experiments have therefore been carried out to probe
the effects of impurities on quantum fluids. For example, scanning
tunnelling microscopy and NMR have examined the response of
high-temperature superconductors to Zn, Ni and Li substituted for
Cu atoms \cite{Hudson,Pan}, while neutron scattering has revealed
that holes create AF droplets with a central $\pi$ phase shift in
the one-dimensional Haldane spin liquid \cite{Xu_Science}.\par

In this Letter we present direct information about vacancy-induced
edge states in the one-dimensional (1D) $S$=$1$ Heisenberg
antiferromagnet (AF) ${\rm Y_2BaNi_{1-x}Mg_xO_5}$ obtained through
a novel inelastic neutron scattering technique. The relative
simplicity of a 1D $S$=$1$ antiferromagnet allows straightforward
interpretation of the data, and therefore can serve as a step
toward understanding impurity-induced phenomena in more
complicated systems. The Hamiltonian for an AF spin chain is
\begin{equation}
    \mathcal{H}=J\sum_i {\bf S}_i\, {\bf S}_{i+1}\, ,
    \label{Hamiltonian}
\end{equation}where $J>0$ is the nearest-neighbor exchange constant.
For $S$=$1$ the ground state of Eq.~\ref{Hamiltonian} is an
isolated singlet in which correlations fall off exponentially and
the excitation spectrum has an energy gap \cite{Haldane83}. The
ground state is similar to that of the Affleck-Kennedy-Leib-Tasaki
(AKLT) model \cite{Affleck_Kennedy}, where the $S$=$1$ states are
described as two $S$=$\frac{1}{2}$ states and the ground state is
made up of local AF valence bonds between $S$=$\frac{1}{2}$ states
of nearest-neighbor sites. The first excited state corresponds to
breaking a valence bond singlet, hence the gap.\par

Non-magnetic impurities in $S$=$1$ chains should produce an
ensemble of finite length chain segments with a singlet-triplet
gap that increases with decreasing chain length \cite{White_Huse}
as well as localized $S$=$\frac{1}{2}$ degrees of freedom at chain
ends \cite{Affleck_Kennedy}. The strongest evidence for
$S$=$\frac{1}{2}$ degrees of freedom from non-magnetic doping was
obtained through ESR \cite{Glarum} and NMR \cite{Tedoldi}
experiments. It is important to note, though, that in real $S$=1
chains, we are dealing with ${\it effective}$ $S$=$\frac{1}{2}$
degrees of freedom, which always occur in $S$=0 and $S$=1 pairs, a
fact graphically illustrated by magnetic specific heat data
\cite{Ramirez}. Here we present inelastic neutron scattering
experiments which probe the microscopic structure of the chain end
states through the wave-vector dependence of the Zeeman
resonance.\par

${\rm Y_2BaNiO_5}$ has a body-centered orthorhombic structure,
space group {\it Immm} with lattice parameters $a=3.76$\,\AA,
$b=5.76$\,\AA\, and $c=11.32$\,\AA $\;$\cite{Buttrey}. The ${\rm
Ni^{2+}}$ chains run along the $a$-axis and the ${\rm Ni^{2+}}$
ions are separated by $\bf{a}$, so ${\bf Q}$=$(h,k,l)$=$(0.5,0,0)$
is the AF point, which is also referred to as the $q=\pi$-point in
theoretical work. The samples were grown by the travelling-solvent
floating-zone method \cite{Ito_Yamaguchi}. The experiments were
performed at the NIST Center for Neutron Research.\par

\begin{figure}[ht]
\begin{center}
  \includegraphics[height=5.5cm,bbllx=66,bblly=260,bburx=500,
  bbury=580,angle=0,clip=]{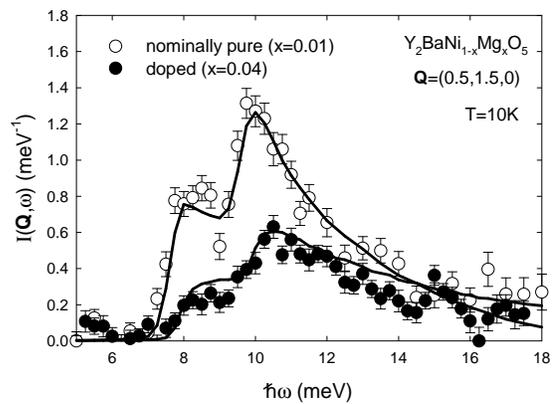}
  \caption{Scattering spectra at $(0.5,1.5,0)$ for doped
  and undoped ${\rm Y_2BaNiO_5}$, shown in absolute units per
  ${\rm Ni^{2+}}$. $I({\bf Q},\omega)$ is the convolution of
  the cross-section with the resolution function. The solid lines
  are the scattering expected for the given distribution of chain
  lengths and convoluted with the resolution function. The nominally
  pure sample, ${\rm Y_2BaNiO_5}$, ($m=0.7\;\mathrm{g}$) was studied
  using BT4 with 40$^{\prime}$-40$^{\prime}$-40$^{\prime}$-80$^{\prime}$
  collimation, a final energy $E_f=14.7\;\mathrm{meV}$ and a PG
  filter in the scattered beam. The experiments on the doped sample,
  ${\rm Y_2BaNi_{0.96}Mg_{0.04}O_5}$, ($m=0.6\;\mathrm{g}$) were
  performed using BT2 with the collimation set to
  60$^{\prime}$-40$^{\prime}$-40$^{\prime}$-80$^{\prime}$, the
  final energy $E_f=13.7\;\mathrm{meV}$ and a PG filter after the
  sample. Intensities were normalized using the integrated intensity of
  a transverse acoustic phonon close to the (200) reciprocal lattice point.}
  \label{Fig1-comp-undop-dop}
\end{center}
\end{figure}

The excitation spectrum of AF $S$=$1$ chains near AF wave-vectors
is dominated by well-defined triplet excitations with an energy
gap $\Delta=0.41050(2)J$ \cite{White_Huse}. Open circles in
Fig.~\ref{Fig1-comp-undop-dop} show the spectrum of the nominally
pure sample, ${\rm Y_2BaNiO_5}$, at the AF point $q=\pi$, measured
at $(0.5,1.5,0)$. A non-magnetic background, determined by
matching constant-${\bf Q}$ scans with the sample offset by
$20^{o}$, was subtracted from the data. Crystal field anisotropies
in ${\rm Y_2BaNiO_5}$ split the Haldane triplet into three
separate modes at $7.5(1)$, $8.6(1)$ and $9.6(1)\;\mathrm{meV}$
for polarizations parallel to the three principal orthorhombic
axes \cite{Xu}. The experiment at $(0.5,1.5,0)$ is least sensitive
to $S^{bb}({\bf Q},\omega)$ because neutron scattering probes the
dynamic spin correlation functions of spin components
perpendicular to the wave-vector transfer ${\bf Q}$:
\begin{equation} \frac{d^2\sigma({\bf Q},\omega)}{d\Omega dE_f}=\frac{k_f}{k_i}(\frac{g}{2}F_{\rm Ni}
({\bf Q}))^2\sum_{\alpha \beta}(1-\hat{Q}_{\alpha}\hat{Q}_{\beta})
S^{\alpha \beta}({\bf Q},\omega)\, .
\end{equation}Here $F_{\rm Ni}({\bf Q})$ is the magnetic form factor
of ${\rm Ni^{2+}}$.\par

Closed circles in Fig.~\ref{Fig1-comp-undop-dop} show the spectrum
for the doped sample, ${\rm Y_2BaNi_{0.96}Mg_{0.04}O_5}$, measured
at the same wave-vector and under similar resolution conditions.
In agreement with work on powder samples of ${\rm
Y_2BaNi_{0.96}Zn_{0.04}O_5}$ \cite{DiTusa}, the spectrum is
attenuated and harder than in the undoped sample, even though the
threshold and peak energies remain essentially the same. This
follows because there is now a distribution of chain lengths with
the Haldane gap rising with decreasing chain length. The longest
chains contribute unchanged spectra and threshold energies, while
the shorter chain spectra are (blue) shifted to higher energies
and produce the enhanced - relative to the peak - high frequency
tail.\par

Indeed the observed neutron spectra follow quantitatively from the
random distribution of the non-magnetic defects in the sample. For
defect density, $x$, the probability that $L+2$ sites form an
isolated chain segment of length $L$ is $P(L,x)=x^2 (1-x)^L$.
Numerical calculations suggest that the energy gap in a chain of
length $L$ is $\Delta^{\alpha}(L)=
\sqrt{(\Delta^{\alpha}_{\infty})^2 +v_s^2 \cdot
\sin^2(\pi(1-1/L))}$ \cite{White_Huse}, where $v_s=2.49(1) J$ is
the spin-wave velocity \cite{Sorensen_3}. The average dynamical
structure factor is then given by $S({\bf Q},\omega)=\sum_L P(L)
S_L({\bf Q},\omega)$, where $S_L({\bf Q},\omega)$ is the structure
factor for a chain segment of length $L$. For simplicity we chose
$S_L({\bf Q},\omega)$ to equal that for an infinite length chain
suitably normalized so as to correspond to $L$ spin and with gap
$\Delta(L)$ \cite{Ma}. $S({\bf Q},\omega)$ was convoluted with the
resolution function \cite{Chesser_Axe} and adjusted to the data in
a global fit. The fit for the nominally pure sample included a
residual impurity concentration, $x=0.010(1)$, that was determined
from the Curie tail in low-$T$ susceptibility data \cite{Xu}. For
the doped sample the impurity concentration was fixed at
$x=0.04$.\par

The only adjustable parameter was a common overall scale factor,
which in the single mode approximation is proportional to the
ground state energy. The Haldane gaps $\Delta^{\alpha}_{\infty}$
and the mode velocity $v_s$ were fixed at the values measured in
the undoped compound \cite{Xu}. This yields $\langle \mathcal{H}
\rangle /L=-1.45(0.4)J$, consistent with numerical results
\cite{White_Huse}. The excellent agreement between experiment and
model in Fig.~\ref{Fig1-comp-undop-dop} validates the assumptions
made above. In particular the model accurately reproduces the
increase of the average gap with impurity concentration $x$ - and
thus provides support for upward renormalization of the gap in
accordance with $\Delta(L)$.\par

We have thus far concentrated on the effects of impurities at
energies near the Haldane gap $\Delta$ of the pure compound. To
introduce what occurs at much lower energies, we recall that
beyond the triplet excitations above $\Delta$, the AKLT model
predicts the existence of a pair of localized composite
$S$=$\frac{1}{2}$ degrees of freedom towards the ends of finite
length $S$=$1$ chains. These are coupled via an indirect exchange
interaction $K(L)=(-1)^L 0.8064 J \exp(-(L-1)/\xi)$
\cite{Sorensen_2}, so that a doped $S$=$1$ chain system can be
described as a set of dimers with a broad distribution of ferro-
and antiferromagnetic interdimer couplings. The ground state of a
chain segment is a singlet when $L$ is even and a triplet when $L$
is odd. For chains that are much longer that the Haldane length
the gap between these levels is much smaller than the Haldane gap.
A magnetic field, $H$, splits the triplet state and inelastic
neutron scattering can then be used to probe transitions between
quantum states of the nanometer-scale chain segments.\par

\begin{figure}[ht]
\begin{center}
  \includegraphics[height=9.0cm,bbllx=70,bblly=125,bburx=515,
  bbury=690,angle=0,clip=]{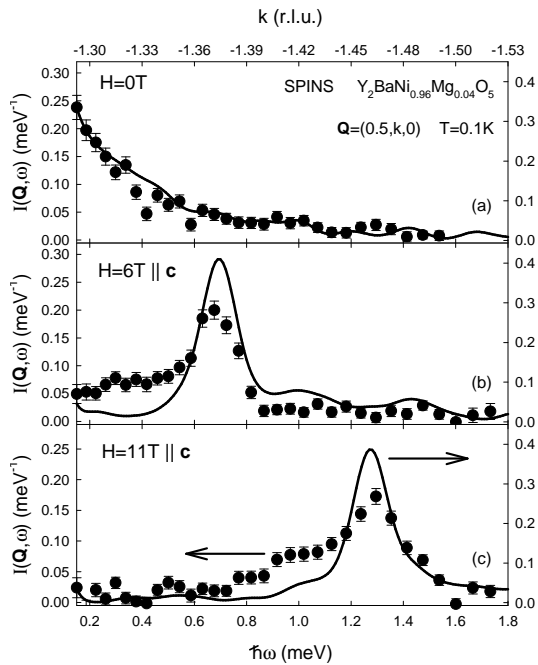}
  \caption{Spectra at ${\bf Q}=(0.5,k,0)$ for three different fields
  along the $c$ axis in absolute units (left axis). The solid line is the
  model described in the text in absolute units (right axis), scaled so
  that the model matches the scattering at zero field. The top scale shows
  the median wave-vector transfer in the transverse direction which was
  varied to maintain ${\bf k_f}$ parallel to the chain direction for
  optimal focusing. The experiment was performed with beam divergences
  40$^{\prime}$-80$^{\prime}$-52$^{\prime}$-300$^{\prime}$ for each of the $11$
  analyzer blade channels, which were set to reflect $E_f=3.7\;\mathrm{meV}$ to
  the center of the detector. A BeO filter rejected neutrons with energies higher
  than $3.7\;\mathrm{meV}$ from the detection system. Normalization of the
  intensity relied on the incoherent elastic scattering from the sample.}
  \label{Fig2-3scans}
\end{center}
\end{figure}

The consideration of the coupled chain end states suggests
searching for excitations with much lower energies than the
Haldane gap. While such excitations cannot exist in pure chains,
they should appear in diluted chains at energies of order the
indirect exchange interactions, and will yield an inelastic tail
extending into the gap from $\hbar\omega=0$, where the chain end
states would reside if the couplings between them were to vanish.
Fig.~\ref{Fig2-3scans}a shows that as expected, this phenomenon
occurs for ${\rm Y_2BaNi_{0.96}Mg_{0.04}O_5}$. To obtain the
spectrum displayed, scattering at $(0.4,k,0)$ and $(0.6,k,0)$ was
averaged and then subtracted as background. For finite fields the
spin fluctuations, including those hidden by elastic incoherent
scattering should be Zeeman-shifted. Our data also agree with this
expectation - the neutron spectra at ${\bf Q}=(0.5,k,0)$
(Fig.~\ref{Fig2-3scans}b and c) show a well-defined excitation,
centered at the Zeeman energy, with clear low energy
sidebands.\par

The data in Fig.~\ref{Fig2-3scans} can largely be explained using
a model of uncoupled dimers consisting of the composite
$S$=$\frac{1}{2}$ degrees of freedom at the chain ends. To include
the spin anisotropies in the system we calculated the neutron
spectra using the low-energy Hamiltonian introduced by Batista
\textit{et al.}\cite{Batista}:
\begin{eqnarray}
{\mathcal H}_L=E^0_L+\left[ J \alpha_L + D\beta_L \right] |0\rangle\langle0| \nonumber\\
+\gamma_L D S^2_z + \gamma_L E (S^2_x + S^2_y)  - \mu_B g_z H_z
S_z\, ,
\end{eqnarray}where $E^0_L$, $\alpha_L$, $\beta_L$, and $\gamma_L$
were computed by the density-matrix renormalization group
technique \cite{Batista} and $|0\rangle\langle0|$ is the
projection onto the singlet state. ${\bf S}$ is the spin operator
for the dimer. The effect of the anisotropy parameters $D=-0.039J$
and $E=-0.0127J$ \cite{Xu} is to split the triplet energy
levels.\par

The neutron scattering cross-section for a dimer is
\begin{eqnarray}
    &\frac{d^2\sigma}{d\Omega dE_f} (L)= (\gamma r_0)^2 (\frac{g}{2}F_{\rm Ni}
    ({\bf Q}))^2 \frac{k_f}{k_i} \sum_{\alpha \beta} \sum_{i,j}
    (\delta_{\alpha\beta}-\hat{\bf Q}_{\alpha} \hat{\bf Q}_{\beta})&
    \nonumber \\ 
    &{F^i_L}^{\star} ({\bf Q}) F^j_L ({\bf Q})\sum_{\lambda_L \lambda_L'}
    p_{\lambda_L} \langle\lambda_L|\exp(-i {\bf Q}  \cdot {\bf R_i})
    S_i^{\alpha}|\lambda_L'\rangle &\nonumber  \\ 
    & \langle\lambda_L'|\exp(i {\bf Q} \cdot {\bf
    R_j})S_j^{\beta}|\lambda_L\rangle
    \delta(E_{\lambda_L}-E_{\lambda_L'}+\hbar\omega)\, ,&
    \label{neutron-cross-section}
\end{eqnarray}where $|\lambda_L\rangle$ denotes the dimer singlet and
triplet states with energy $E_{\lambda_L}$ and population factor
$p_{\lambda_L}$. The index $i,j=1,2$ numbers the composite
$S$=$\frac{1}{2}$ at a position $R_i$ towards the end of the chain
and $\alpha,\beta=x,y,z$. We used $g=g_z=2.165$ as determined from
ESR \cite{Batista} and $(\gamma r_0)^2=0.29\;\mathrm{barn}$.
$F^i_L({\bf Q})$ is the form factor of the end-chain composite
spin and describes the decaying AF magnetization at the chain ends
associated with the collective spin degree of freedom. We will
give an experimental view of this quantity below; for the moment,
we simply assert that its square is well described by a normalized
Lorentzian centered at $h=0.5$ with a width $\Gamma$ given by
$\Gamma^2=(\frac{1}{\xi^2}+\frac{1}{L^2})/\pi^2$. This should be a
good approximation because the Fourier transform of an
exponentially decaying correlation function for the staggered
magnetization is a Lorentzian, and the finite length $L$ of the
chain acts as a finite size cutoff.\par

The scattering cross-section for a given dimer distribution was
obtained by adding contributions from all chain segment lengths
$\frac{d^2\sigma}{d\Omega dE}= \sum_L P(L,x)
\frac{d^2\sigma}{d\Omega dE}(L)$ and convoluting with a Gaussian
to account for the finite energy resolution. Lacking detailed
singlet and triplet wave functions for a chain segment,
interference effects, such as the twin incommensurate peaks for
the subgap bound states in hole-doped ${\rm Y_{\rm
2-x}Ca_{x}BaNiO_{5}}$ \cite{Xu_Science}, between chain ends were
ignored. This should be a good approximation for an ensemble of
long chains. The solid line in Fig.~\ref{Fig2-3scans} shows the
absolute scattering intensity expected from this model. The weakly
oscillatory nature of the calculated spectrum follows from the
discrete nature of the distribution of indirect exchange couplings
$K(L)$. At zero field there is excellent agreement between
experiment and theory with an $28\%$ adjustment of the pre-factor
(within the $30\%$ error bar from normalization) being the only
adjustable parameter. At finite field the model reproduces the
position of the well-defined peak ($g_z=2.0(0.05)$ gives a
slightly better fit than $g_z=2.165$ from ESR \cite{Batista}) but
it overestimates its intensity, which seems to be transferred to a
continuum that cannot be accounted for by the model. The broad
scattering below the well-defined excitation indicates that
changes in the chain end structure with field or interchain
coupling play an important role. If interchain interactions indeed
couple the chain segments then the doped sample may eventually
freeze into a low-temperature AF or spin glass state.\par

\begin{figure}[ht]
\begin{center}
  \includegraphics[height=5.6cm,bbllx=65,bblly=260,bburx=500,
  bbury=570,angle=0,clip=]{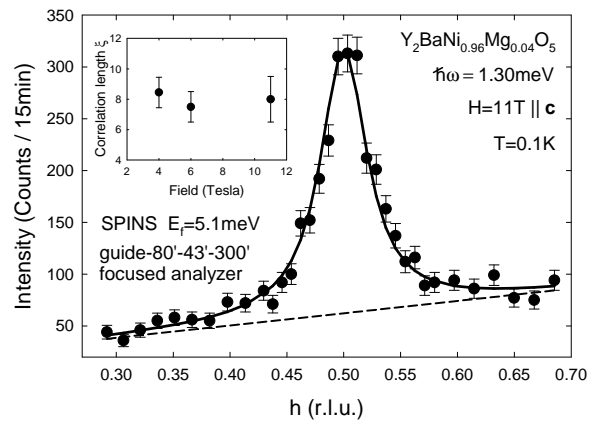}
  \caption{Scattering originating from $S$=$\frac{1}{2}$ chain end
  states at the resonance energy $\hbar\omega=1.30\;\mathrm{meV}$ for
  $H=11\;\mathrm{T}$ (applied along the $c$ axis) in
  ${\rm Y_2BaNi_{0.96}Mg_{0.04}O_5}$ as a function of $h$. The solid
  line is a fit explained in the text and the dashed line is the
  background. Inset: Correlation length $\xi$ as a function of field.
  For this experiment $E_f=5\;\mathrm{meV}$ and there was a cooled Be
  filter in the scattered beam.}
  \label{Fig3-SQa}
\end{center}
\end{figure}

Neutron scattering has the tremendous advantage of providing
spatial as well as spectral information about spin fluctuations.
We have thus examined the form factor for the edge states at the
Zeeman energy by scanning wave vector $h$ with fixed
$\hbar\omega$. Fig.~\ref{Fig3-SQa} shows the result. Rather than a
broad maximum centered at $h=0$ corresponding to the form factor
for a localized atomic spin, we find a relatively sharp peak
centered at $h=0.5$. This is direct evidence that chain end spins
carry AF spin polarization back into the bulk of the chain
segments. By fitting the model cross-section
(Eq.~\ref{neutron-cross-section}) convoluted with the resolution
function to the data we find the exponential decay length $\xi$
for the chain-end staggered spin density (inset of
Fig.~\ref{Fig3-SQa}). The decay length remains constant at
$\xi=8(1)$ between $H=4$ and $11\;\mathrm{T}$ - in perfect
agreement with the equal-time correlation length $\xi=8(1)$ for
undoped ${\rm Y_2BaNiO_5}$ in zero field at low temperatures $T/J
\ll 1$ \cite{Chen_Nature}, but slightly longer than that deduced
from an NMR study of the staggered magnetization \cite{Tedoldi}. A
fit to a square-root Lorentzian, which would be appropriate for
unpinned (by impurities) bulk spin correlations from the Haldane
continuum, yields a longer correlation length $\xi=13(1)$. Note
that because we have fixed $\hbar\omega$ at precisely the Zeeman
energy, we have maximized the contribution of those chains with
the smallest coupling and hence largest lengths to the scattering,
thus revealing $F^{i}_{L\simeq\infty}({\bf Q})$, in the notation
of Eq.~\ref{neutron-cross-section}.\par

In summary, our inelastic neutron scattering experiment on
Mg-doped ${\rm Y_2BaNiO_5}$ shows an upward renormalization of the
Haldane gap with doping, which is consistent with numerical
predictions for a distribution of finite chain segments. In
addition, we find well-defined excitations below the Haldane gap
whose field dependence demonstrates that they originate from
$S$=$\frac{1}{2}$ chain-end degrees of freedom. Their wave-vector
dependence shows chain end states carry an AF "Friedel"
oscillation that extends into the chain segment with an
exponential decay length $\xi=8(1)$ in a field $H=11\;\mathrm{T}$.
The experiment also demonstrates a new technique, neutron ESR, by
which the structure of composite spin degrees of freedom liberated
from a macroscopic singlet can be probed. Efforts are now underway
to apply this approach to understand impurity structure and
dynamics in more complex systems such as high-temperature
superconductors \cite{Lake}.

\begin{acknowledgments}
We thank C.~D. Batista for sending us numerical results. Work at
JHU was supported by the NSF through Grant No. DMR-0074571. Work
at LSU was supported through Grant No. DMR-0103892. Work at BNL
was supported by DOE Division of Materials Sciences under Contract
DE-AC02-98CH10086. The work at SPINS was supported by NSF through
DMR-9986442. The high-field magnet used was funded in part by NSF
through DMR-9704257.
\end{acknowledgments}


\begin{thebibliography}{10}

\bibitem{Pan}
S.~H. Pan \textit{et al.} Nature {\bf 403},  746  (2000).

\bibitem{Hudson}
E.~W. Hudson \textit{et al.} Nature {\bf 411},  920  (2001).

\bibitem{Xu_Science}
G.~Y. Xu \textit{et al.} Science {\bf 289},  419 (2000).

\bibitem{Haldane83}
F.~D.~M. Haldane, Phys. Rev. Lett. {\bf 50},  1153  (1983).

\bibitem{Affleck_Kennedy}
I. Affleck \textit{et al.} Phys. Rev. Lett. {\bf 59}, 799  (1987).

\bibitem{White_Huse}
S.~R. White \textit{et al.} Phys. Rev. B {\bf 48},  3844  (1993).

\bibitem{Glarum}
S.~H. Glarum \textit{et al.} Phys. Rev. Lett. {\bf 67},  1614
(1991).

\bibitem{Tedoldi}
F. Tedoldi \textit{et al.} Phys. Rev. Lett. {\bf 83},  412 (1999).

\bibitem{Ramirez}
A.~P. Ramirez \textit{et al.} Phys. Rev. Lett. {\bf 72}, 3108
(1994).

\bibitem{Buttrey}
D.~J. Buttrey \textit{et al.} J. Solid State Chem. {\bf 88},  291
(1990).

\bibitem{Ito_Yamaguchi}
T. Ito \textit{et al.} Phys. Rev. B {\bf 64},  060401(R)  (2001).

\bibitem{Xu}
G. Xu \textit{et al.} Phys. Rev. B {\bf 54},  R6827  (1996).

\bibitem{DiTusa}
J.~F. DiTusa \textit{et al.}, Phys. Rev. Lett. {\bf 73},  1857
(1994).

\bibitem{Sorensen_3}
E.~S. Sorensen \textit{et al.} Phys. Rev. Lett. {\bf 71},  1633
(1993).

\bibitem{Ma}
S. Ma \textit{et al.} Phys. Rev. Lett. {\bf 69},  3571  (1992).

\bibitem{Chesser_Axe}
N.~J. Chesser and J.~D. Axe, Acta Crys. {\bf A29},  160  (1973).

\bibitem{Sorensen_2}
E.~S. Sorensen \textit{et al.} Phys. Rev. B {\bf 49},  15771
(1994).

\bibitem{Batista}
C.~D. Batista \textit{et al.}, Phys. Rev. B {\bf 60},  R12553
  (1999).

\bibitem{Chen_Nature}
Y. Chen \textit{et al.}, to be published.

\bibitem{Lake}
B. Lake \textit{et al.}, Science {\bf 291},  1759  (2001).

\end{thebibliography}
\end{document}